\documentclass[prl,twocolumn]{revtex4}
\begin{document}

\title{ Modular values  and weak values of quantum observables. }

\author{Y. Kedem and L. Vaidman}
\affiliation{ Raymond and Beverly Sackler School of Physics and Astronomy\\
 Tel-Aviv University, Tel-Aviv 69978, Israel}

\begin{abstract}
The concept of a modular value of an observable of a pre- and post-selected quantum system is introduced. It is similar in form and in some cases has a close connection to the weak value of an observable, but instead of describing an effective interaction when the coupling is weak, it describes a coupling of any strength  but only to qubit meters. The generalization of the concept for a coupling of a composite system to a multi-qubit meter provides an explanation of some current experiments.
  \end{abstract}
%\pacs{03.65.Ta, 03.65.Ud, 03.67.Bg, 42.50.Xa}
\maketitle

In 1988 Aharonov, Albert and Vaidman (AAV) \cite{AAV88} discovered that a weak coupling to any observable $C$ of a pre- and post-selected system becomes an effective coupling to a ``weak value" of this observable.  Recently, there has been increased interest in weak values, in particular due to the successful applications of the AAV effect as an amplification scheme for high precision measurements of tiny effects \cite{HK,How}. In 2004  Resh and Steinberg considered the weak values of nonlocal observables in composite systems \cite{RS04}, culminating in experimental demonstrations \cite{Ha-ex1,Ha-ex2} of the Hardy paradox \cite{Ha,A-ha}. While most research on weak values considered coupling to  continuous variables, increasing  attention has turned to weak values appearing in couplings to qubit meters \cite{Brun,Mo}.

 In this letter we introduce a new concept, the {\it modular value of an observable}  which characterizes the  coupling of pre- and post-selected quantum systems to qubit variables. The  modular values can be measured efficiently using strong coupling.
  In some cases, modular values are equal to weak values. Measurements of modular values, then, provide  an efficient  method of measuring these weak values. This is particularly important for measuring weak values of nonlocal observables.

In the two-state vector formalism, a quantum system at  times $t$, intermediate between  two measurements, is described by a
 {\em two-state vector} \cite{AV90}:
\begin{equation}\label{tsv}
   \langle \phi |~~|\psi\rangle ,
\end{equation}
where $|\psi\rangle$ is the state prepared before $t$ and the state  $| \phi \rangle$ was found after $t$. For simplicity we assume  that the free Hamiltonian is zero.

 The most general statement of the AAV effect is that weak coupling at the intermediate times $t$  to any physical observable $C$ of the system is an effective coupling to the {\it weak value} of $C$:
\begin{equation}\label{wv}
C_w \equiv { \langle{\phi} \vert C \vert\psi\rangle \over
\langle{\phi}\vert{\psi}\rangle } .
\end{equation}
The operational meaning of this is that  during the time between the two strong measurements, when  considering the action of our system on other systems,  we can  replace any operator  of our system by the corresponding  $c$-number:
 \begin{equation}\label{repla}
    C ~\rightarrow ~C_w .
\end{equation}

In particular, coupling the observable $C$ to a continuous variable $P$ (as in the von Neumann measurement procedure), via
\begin{equation}\label{neumann}
   H = g(t) PC ,
\end{equation}
with
\begin{equation}\label{norm}
  \int g(t) dt =k,
\end{equation}
such that  the  interaction Hamiltonian (\ref{neumann}) is weak, leads to an effective Hamiltonian of the external system (measuring device)
\begin{equation}\label{wc}
   H = g(t)C_w P .
\end{equation}
For the standard choice, $k=1$, the interaction leads to a ``shift'' of the wave function of the conjugate variable $Q$:
\begin{equation}\label{tiev0}
   \Psi_{fin} (Q) ={\cal N} \Psi (Q-C_w) ,
\end{equation}
where ${\cal N}= (\int |\Psi (Q-C_w)|^2 dQ)^{-{1\over 2}}$ is the normalization factor (which is equal 1 when $C_w$ is real).

For real $C_w$, the shift of the probability distribution of the (pointer) variable $Q$ (the outcome of the von Neumann procedure) is equal to the corresponding weak value. Note that the shift is the only effect of the measurement interaction, the shape of  the distribution remains unchanged.

If the weak value is a complex number, the effective Hamiltonian is non-Hermitian, which is a less familiar situation \cite{Adiab}.
In the context of weak measurements, the simplest picture is obtained when the state of the measurement system is given by a Gaussian  $ \Psi_{in} (Q)  =(\Delta ^2 \pi )^{-1/4}  e^{ -{{Q ^2} /{2\Delta ^2}}} $. Then, the shift of the distribution of the pointer variable  provides the real part of the weak value:
\begin{equation}\label{tiev}
   \rho_{fin}(Q)=\Psi^\ast_{fin}  \Psi_{fin}={1\over{\Delta \sqrt{\pi }}} e^{ -{{(Q - {\text Re} C_w)^2} /{\Delta ^2}}},
\end{equation}
while the shift of the Gaussian in the $P$ representation provides the imaginary part:
\begin{equation}\label{tiev2}
\tilde \rho_{fin}(P)= \tilde \Psi^\ast_{fin}  \tilde  \Psi_{fin} ={\Delta \over \sqrt{\pi }} e^{ -{(P - {\text {Im}} C_w / \Delta^2)^2 \Delta ^2}}.
\end{equation}
Note that measuring ${\text {Im}} C_w$ is particularly helpful for practical applications of the AAV effect \cite{Brunner}.

The formal shift (\ref{tiev0}) is  valid for any wave function of an external system, but it is not observable directly because $Q$ is real. Simple shifts (without distortion) of the distributions  of $Q$ and $P$, (\ref{tiev}),(\ref{tiev2}) are special features of a Gaussian.
For a complex valued wave function, the two distributions  depend on both the real and the imaginary parts of the weak value \cite{Joz} and are better described using cumulants \cite{Mich}.

There are two different physical methods  which transform  standard von Neumann measurement  procedure to a weak measurement. The first, described above, is to keep the coupling the same as in a strong measurement, but to change the initial state of the measuring device. The second, is to keep the initial state of the measuring device, but reduce the strength of the coupling characterized by $k$ in Eq.(\ref{norm}).

 The AAV effect is not limited to couplings to continuous variables as in the von Neumann measurement procedure. If an observable $C$ of a pre- and post-selected system is coupled to a discrete variable, the replacement (\ref{repla}) will still lead to a correct effective Hamiltonian for weak enough coupling \cite{Mo}. In this case, however, we do not have two options for enforcing  weakness of the interaction. For discrete variables,  we  may not achieve weakness  by preparing a particular initial state, but rather we have to reduce the strength of the coupling.
For example,  in the coupling to a spin-$1\over 2$ particle:
\begin{equation}\label{wcSpin}
   H =g(t) \sigma_z C ,
\end{equation}
$\sigma_z$ cannot be made small. Instead, we reduce  $g(t)$ such that  $ k \ll 1$, and this does validate  the replacement (\ref{repla}). Then, if the initial state of the spin is
 $
 \alpha \left|\uparrow\right\rangle +\beta\left|\downarrow\right\rangle
$,
the final state of the spin is
 \begin{equation}\label{finstate1}
     {\cal N}( \alpha e^{-ikC_w} \left|\uparrow\right\rangle +\beta e^{ikC_w} \left|\downarrow\right\rangle).
\end{equation}
Thus, the effect of the weak coupling is a small change in the direction of the spin: $\delta \phi = 2 k \, {\text {Re}} C_w$  and  $\delta \theta =- 2 k \, \sin{ \theta} \, {\text {Im}} C_w  $.

Let us consider another  coupling to a qubit:
\begin{equation}\label{mvHam}
   H =g(t) \textbf{P} C ,
\end{equation}
where the qubit has  states $|0\rangle$ and  $|1\rangle$, and  $\textbf{P}\equiv\left|1\right\rangle \langle 1 |$.
Let the initial state of the qubit  be
 \begin{equation}\label{instate1}
 \alpha |0 \rangle +\beta |1\rangle .
\end{equation}
Here, the choice $\beta \ll 1$ in  the initial  state, enforces small expectation value of the interaction Hamiltonian (\ref{mvHam})
 while keeping  the coupling strong, $k=1$.
 However, the substitution (\ref{repla})
which leads to the final state
\begin{equation}\label{finstate2}
   {\cal N}(\alpha \left|0\right\rangle +\beta e^{-ikC_w}\left|1\right\rangle) ,
\end{equation}
does not provide the correct description in this case. The correct analysis of this example will lead to the main result of our letter.

The  state of the qubit, after the interaction (\ref{mvHam}) and post-selection of the state of the system $|\phi\rangle$, is
   \begin{equation}\label{afterevol}
 {\cal N}(\alpha |0\rangle +\beta{{\langle \phi |~e^{-ikC}~|\psi\rangle} \over {\langle \phi |\psi\rangle}} |1\rangle) .
\end{equation}
(In general, this is  different from (\ref{finstate2}), but the two expressions are equal in the limit of weak coupling $k \ll 1$.)
The action of the pre- and post-selected system on the qubit is completely described by a single complex number which we name {\it modular value}:
   \begin{equation}\label{mv}
 C_m\equiv{{\langle \phi |~e^{-ikC}~|\psi\rangle} \over {\langle \phi |\psi\rangle}} .
\end{equation}
The modular value can be found through a measurement on an ensemble, by making tomography measurement of the final state of the qubit. If the final state is found to be
$
 \gamma |0\rangle +\delta|1\rangle ,
$
then, comparing  with (\ref{afterevol}), we obtain:
\begin{equation}\label{mvcal}
 C_m ={ {\alpha \delta} \over{\beta\gamma}}.
\end{equation}

The modular value  resembles a weak value in its form. It has the same ``amplification'' factor $1\over{\langle \phi |\psi\rangle}$. Further, just like the weak value, it is defined only for post-selected states which are not orthogonal to the pre-selected state. However, the reasons for this are different. The weak value  describes a weak coupling of a pre- and post-selected quantum system. In the limit of vanishing coupling, the post-selection of an orthogonal state is impossible. The modular value is defined for a particular finite coupling strength  $k$. Thus,  when it is measured at an intermediate time, the post-selection of an orthogonal state is, in general, possible. However, in this case, the final state of the qubit meter is always $|1\rangle$ and thus, the only information we can learn from tomography of the final state is that
$\langle \phi |~e^{-ikC}~|\psi\rangle\neq0$.

Consider the modular value  of a spin-$1\over2$  with $k={\pi\over2}$:
  \begin{equation}\label{mv=wv}
 \sigma_m ={{\langle \phi |~e^{-i{\pi\over 2} \sigma} ~|\psi\rangle} \over {\langle \phi |\psi\rangle}}= {{\langle \phi |-i\sigma~|\psi\rangle} \over {\langle \phi |\psi\rangle}}=-i\sigma_w.
\end{equation}
The modular value of a spin component yields its weak value, but the former can be measured using a coupling which is not weak. Thus, weak values appear not only in the AAV effect, i.e., not only as effective interactions in the limit of the weak coupling.

Another situation in which we can see a manifestation of weak values outside the the range of validity of the AAV effect is the weak values of nonlocal observables, such as the product of variables $A$ and $B$ related to separate parts of a composite system. Formally, we can consider  $(AB)_w$, but it will have no meaning as an effective interaction. Indeed, replacing an observable in the Hamiltonian by the $c$-number weak value requires that the observable will appear in the Hamiltonian in the first place. Relativistic quantum mechanics puts constraints on the allowed interactions, and products $AB$ where $A$ and $B$ are observables related to separate locations in space, cannot appear in the Hamiltonian.

Surprisingly, Resch and Steinberg (RS) \cite{RS04} showed that it is still possible to find $(AB)_w$ from a statistical analysis of correlations of the outcomes of local measuring devices (\ref{neumann}), one interacting with $A$ and the other with $B$.
Starting with the initial  product state  $ \Psi_{in} (Q_A,Q_B)  =(\Delta ^2 \pi )^{-1/2}  e^{ -{{(Q_A ^2+Q_B ^2)} /{2\Delta ^2}}} $, they find  \cite{RL}:
\begin{equation}
{\text {Re}}(AB)_{w}={1 \over k^2} \left( \langle Q_{A}Q_{B}\rangle - 4 \Delta^4 \langle P_{A} P_{B}\rangle \right). \label{laudeen}
\end{equation}
The RS method is universal: it is applicable to any local observables $A$ and $B$ and it was generalized to a product of any finite number of local observables. The analysis of correlations of local measurements of observables $C_i$ (each related to location $i$) allows calculation of any product $\prod_{i=1}^N C_i$ \cite{Resch}. The drawback of the method is that it requires a very large ensemble, since it is based on an $N$th order effect \cite{BV}.
We will show below that for a particular case of product of qubit variables we can find the weak values much more efficiently by measuring the modular values instead.

Let us show now that the modular value of the sum of a set of observables $C_i$ related to various locations $i$ is measurable using qubit meters placed at each location. To this end, we use interaction Hamiltonian
\begin{equation}\label{mvHam1}
   H =g(t) \sum \textbf{P}_i C_i  ,
\end{equation}
where $\textbf{P}_i=\left|1\right\rangle_i \langle 1 |_i$ is the projection on the state of  qubit $i$ and $g(t)$ defines  the coupling strength $k$.
If the initial state of the multiple qubit measuring device is
 \begin{equation}\label{instateN}
 \alpha \prod|0\rangle_i +\beta\prod|1\rangle_i ,
\end{equation}
 then the final state is
  \begin{equation}\label{finstateNN}
 \alpha \prod |0\rangle_i +\beta \left(\sum C_i\right )_m\prod|1\rangle_i .
\end{equation}
The tomography of the final state performed on an ensemble yields the modular value of the  sum $(\sum C_i)_m$.

Minor modifications of this method will allow us to measure modular values of any partial sum $\sum_{i\in\Omega}C_i$, by choosing the initial state
\begin{equation}\label{finstateNmod}
 \alpha \prod|0\rangle_i +\beta\prod_{i\in\Omega}|1\rangle_i\prod_{i\notin \Omega}\left|0\right\rangle_i,
\end{equation}
or any linear combination of $C_i$, by choosing different strengths $k_i$ for the local couplings.

For qubit variables, modular value of a sum with $k={\pi\over 2}$ yields  the weak value of a product:
\begin{equation}\label{summv=prowv}
 \left(\sum_{i=1}^N\sigma_i \right)_m =(-i)^N\left(\prod_{i=1}^N \sigma_i\right)_w.
\end{equation}
For $N$ spins, measuring the weak value of the product using our method  requires tomography of the $N$-qubit state (\ref{finstateNN}), but  still is much easier to perform than the measurements in the RS approach based on the observation of correlations of $N$ Guassian-state pointers \cite{Resch}.

Let us apply our method to the analysis of the Hardy paradox \cite{Ha} which has recently been extensively analyzed theoretically and experimentally  \cite{A-ha,Ha-ex1,Ha-ex2}. In Hardy's setup there are two pre- and post-selected particles  at four separate locations $A,B,C,D,$ described  by a two-state vector:
 \begin{equation}\label{hardy}
  {1\over2}\!\left(\langle A |\!-\! \langle C | \right) \left(\langle B  |\!- \!\langle D | \right)  ~~ {1\over\sqrt 3} \! \left( |A \rangle |D\rangle\! +\!|C \rangle |D\rangle\! +\!|C \rangle |B\rangle \right).
\end{equation}
The paradoxical property tested using weak measurements is that the product rule does not hold:
 \begin{equation}\label{A-B-AB}
 (\textbf{P}_A^1)_w=1, (\textbf{P}_B^2)_w=1,~~~~  \text{but}~~~(\textbf{P}_A^1 \textbf{P}_B^2)_w=0.
\end{equation}

A similar paradox can be formulated for two pre- and post-selected spin-$1\over 2$ particles described  by a two-state vector  \cite{V-PRLHa}:
  \begin{equation}\label{tsvsxsy}
 \langle {\uparrow_y} {\uparrow_x} | ~~~~  {1\over \sqrt{2}}\left( |{\uparrow_z} {\downarrow_z}\rangle - |{\downarrow_z} {\uparrow_z}\rangle  \right),
\end{equation}
where $\langle {\uparrow_y} {\uparrow_x} | =\langle {\uparrow_y}|_1 \langle{\uparrow_x} |_2$, $|{\uparrow_z} {\downarrow_z}\rangle =|{\uparrow_z}\rangle_1 |{\downarrow_z}\rangle_2$ etc. The  failure of the product rule here is seen as follows:
 \begin{equation}\label{qubitprod2}
(\sigma_x^1)_w=-1, (\sigma_y^2)_w=-1,~~~~ \text{but}~~~(\sigma_x^1 \sigma_y^2)_w=-1 .
\end{equation}
Using  (\ref{summv=prowv}), the results (\ref{qubitprod2}) can be demonstrated directly using measurements of the modular values $(\sigma_x^1)_m$, $(\sigma_y^2)_m$, and $(\sigma_x^1 + \sigma_y^2)_m$.

For the measurement of modular values of several observables there are  two conceptually different methods. In the first method, we split the pre- and post-selected ensemble and measure on each sub ensemble a modular value of a single observable, such that each member of the ensemble is coupled to a single meter.  Another approach is a ``weak'' regime of modular values measurement in which we choose the initial states of the qubit and multi-qubit meters (\ref{instate1}), (\ref{instateN}), respectively,   with  $\beta \ll 1$.  We can then  couple  each member of the ensemble with the meters of all observables without causing significant disturbance.
(Note again that limit $\beta \ll 1$ does not make the replacement (\ref{repla})  valid.)

The original Hardy setup has been experimentally realized by two groups. Both used  polarization variables of pairs of photons. In the first experiment \cite{Ha-ex1}, the polarization variables were not entangled, the coupling was weak, and the analysis followed the RS approach \cite{RS04}. In the second experiment, the entangled photons flip polarization  in the interaction region \cite{Ha-ex2}. We will argue that the second experiment was not a weak measurement, but a measurement of  modular values, which, surprisingly, yielded weak values.

The two qubits of the measuring device were the  polarization variables of the photons themselves. The local interactions caused vertical polarization state to acquire the phase of $\pi$: $|H\rangle \rightarrow |H\rangle, ~~ |V\rangle \rightarrow  -|V\rangle$. This corresponds to the interaction Hamiltonian:
 \begin{equation}\label{mvHam2}
   H =g(t)\left (\textbf{P}_V^1~\textbf{P}_A^1 +  \textbf{P}_V^2~\textbf{P}_B^2\right ),
\end{equation}
where $\textbf{P}_V^{i}= |V\rangle_{i} \langle V|_{i}$,  $\textbf{P}_A^1 = |A\rangle_1 \langle A|_1$, $\textbf{P}_B^2 = |B\rangle_2 \langle B|_2$, and $g(t)$ such that $k=\pi$. This is the Hamiltonian for measuring  $(\textbf{P}_A^1 + \textbf{P}_B^2)_m$ as well as $(\textbf{P}_A^1)_m$ and $(\textbf{P}_B^2)_m$ separately, depending on the type of the initial polarization states,  (\ref{instate1}) or (\ref{instateN}).

In the Hardy experiment we  are interested in  measuring  the weak values of the projections (\ref{A-B-AB}). But we  can  express them through modular values, $k=\pi$. Indeed, simple calculations, based on the identity $ e^{-i\pi \textbf{P}}=1-2\textbf{P}$ show the following identities:
\begin{eqnarray}\label{weak_mod}
  (\textbf{P}_A^1)_w \! = \! {1 \over 2} \left( 1\!-\!(\textbf{P}_A^1)_m\right),~  (\textbf{P}_B^2)_w \!= \!{1 \over 2} \left( 1\!-\!(\textbf{P}_B^2)_m\right),~ \\
  (\textbf{P}_A^1  \textbf{P}_B^2)_w\! =\!  {1 \over 4} \left( (\textbf{P}_A^1\! +\! \textbf{P}_B^2)_m\! - \!(\textbf{P}_A^1)_m \!-\! ( \textbf{P}_B^2)_m\! +\! 1\right).~
\end{eqnarray}
Thus, measuring modular values yield weak values.

In the actual experiment \cite{Ha-ex2}, instead of preparing  the initial states of the qubits in the form (\ref{instate1}) or (\ref{instateN}), all pairs of qubits were prepared in the  state:
\begin{equation}\label{instate4}
 {1 \over \sqrt{1+3 |\beta|^2}} \left(|00\rangle +\beta (  |01\rangle +  |10\rangle +|11\rangle)\right ),
\end{equation}
where $|00\rangle = |H\rangle_1  |H\rangle_2$,  $|01\rangle = |H\rangle_1  |V\rangle_2$, etc. This initial state corresponds to a ``superposition'' of measurements of modular values of several observables: the wave function of the qubits is changed as the function of all these modular values. Indeed, the final state is
\begin{equation}\label{finstate4}
 {\cal N}\! \left(|00\rangle \! +\! \beta ( (\textbf{P}_B^2)_m |01\rangle \! + \!(\textbf{P}_A^1)_m |10\rangle \! + \!(\textbf{P}_A^1 \! + \! \textbf{P}_B^2)_m|11\rangle)\!\right ).
\end{equation}
Only partial  tomography  of this final state was performed: the qubit measurements were performed  in a single basis, ${1\over\sqrt 2}(|0\rangle_1\pm|1\rangle_1)~{1\over\sqrt 2}(|0\rangle_2\pm|1\rangle_2)$. For small $\beta$, these measurements provided a good evaluation of the real part of weak values (\ref{A-B-AB}).  It is not surprising that this method was more efficient for measurement of nonlocal weak values than experiment \cite{Ha-ex1} which was based on the second order effect of local weak measurements (the RS method). With  proper tomography, this experiment could have been performed for large   $\beta$, showing both real and imaginary parts of the weak values with even much better efficiency.

The modular value of an observable is a property of a pre- and post-selected quantum system which provides a complete descriptions of how it affects a qubit via general interaction of the form (\ref{mvHam}).  This concept explains the appearance of weak values in quantum-gate type interactions \cite{Brun,Ha-ex2}.
In particular, it provides a scalable method for measuring weak values of products of qubit variables of composite systems, where the direct method is very inefficient.
Due to simplicity of its form (\ref{mv}), modular values  have a potential for useful applications in devising and engineering  novel quantum  protocols.

We should mention two very recent works which might seem related. In the first \cite{Jeff}, modular variables, weak values, and post-selection were applied for explaining  the nature of quantum interference through non-local equations of motion of (non-local) modular variables. The basic difference here is that in this work the system was coupled to a continuous variable and ``modularity'' was the property of the system. In the second work \cite{DAJ}, the concept of ``contextual values'' yielded weak values in some cases, but the context in which the weak values were obtained corresponded to the AAV effect.

This work has been supported in part by the Binational Science Foundation Grant No. 32/08,
 the Israel Science Foundation  Grant No. 1125/10, and  The Wolfson Family Charitable Trust.

\end{document}